# The MICADO first light imager for ELT: its astrometric performance

J.-U. Pott[*a], G. Rodeghiero[a], H. Riechert[a], D. Massari[b], M. Fabricius[c], C. Arcidiacono[d], R. I. Davies[c]

[a] Max-Planck-Institute for Astronomy (MPIA), Heidelberg, Germany; [b] Univ. Of Groningen, NL; [c] Max-Planck-Institute for Extraterrestrial Physics (MPE), DE; [d] INAF-Bologna, IT

## ABSTRACT

We report on our ongoing efforts to ensure that the MICADO NIR imager reaches differential absolute (often abbreviated: relative) astrometric performance limited by the SNR of typical observations. The exceptional 39m diameter collecting area in combination with a powerful multi-conjugate adaptive optics system (called MAORY) brings the nominal centroiding error, which scales as FWHM/SNR, down to a few 10 µas. Here we show that an exceptional effort is needed to provide a system which delivers adequate and calibrateable astrometric performance over the full field of view (up to 53 arcsec diameter).

**Keywords:** ELT, near-infrared, adaptive optics, astrometry, distortion calibration,

## 1. INTRODUCTION

In the contribution, we will present our progress on the understanding of the specific error budget of the astrometric performance of the ELT/MICADO first light imager. Despite of being challenging, high-performance astrometry with a ground-based imager behind an ELT scale aperture is particularly tempting, since the astrometric efficiency (astrometric precision / source brightness / time) increases with $D^3$ for a field of background limited point sources. Since astrometric projects often target proper motions, and require repeated observations, any increase in astrometric efficiency offers high return-on-investment.

MICADO will be the first imager behind a wide-field multi-conjugate adaptive optics system, which is designed from the start to reach field-differential (often called *relative*) astrometric performance at ≤50 micro-arcsec (µas) levels over the full field of up to 53" diameter, delivered by a 3x3 array of Hawaii4RG detectors at 4mas/pix scale. While the excellent 10mas diffraction limited resolution at 1.6 microns of the 39m aperture is attractive, the realization via a 5 Mirror ELT, a complex multi-DM MCAO system with deformable mirrors far from the pupil, and eventually a throughput optimized imager with large opto-mechanics bears several challenges. MICADO will go through its preliminary design (phase B) review process in fall 2018, so SPIE is timely to discuss our current understanding of the interplay of on-sky, and day-time calibrations of the hardware, as well as specific astrometry relevant specifications, such as mid-spatial frequency errors of surface figures, plate-scale and rotation control during an astrometric exposure, chromatic effects of atmosphere and optics, and the alignment stability of the nominal optical design. We will also discuss a spatial scale dependence of the actual astrometric measurement precision to be done with MICADO, which focuses on observing the proper motion kinematics in dense stellar systems (stellar clusters, galactic nuclei, dwarf galaxies), as opposed to global astrometric missions (Gaia, VLBI).

## 2. ORDER-OF-MAGNITUDE CONSIDERATIONS

The astrometric calibration challenge is a two-step effort:

1. Distortion correction: Remove the spatial variance of the plate scale over the entire field-of-view (FoV), so that a given angular distance (separation between two stars), measured at various locations and orientations in the FoV appears the same.
2. Calibrate the global plate scale over the FoV

---

[*] jpott@mpia.de

Can (1) and (2) be done at the 10 µas levels with a straight forward calibration approach? MICADO is a multi-purpose first light near-infrared imager and spectrograph for a not yet built telescope, and MCAO-supported astrometry is only one of its envisioned scientific capabilities among others. For these reasons, we decided to explore the limitations of a distortion calibration without excessive real-time metrology and calibration needs throughout the system.

While the diffraction- and signal-to-noise (SNR)-limited centroiding precision of MICADO reaches the 10 µas level relatively easily (in a few 10sec of integration time on a 16mag star), ensuring a comparable plate scale accuracy of 10 µas over >10arcsec field in the instrument, everywhere and at the same time, would mean unmatched distortion calibration at the $10^{-6}$ level or better, which is clearly beyond our goal.

To put this in context: The most recent ground-based high-performance astrometry mission in the near-infrared, the VLTI-Gravity instrument, just went on sky (Eisenhauer et al. 2011) to deliver 10 µas over 1arcsec distances for a single pair of stars (ie. plate scale accuracy $10^{-5}$). As VLTI 2nd generation instrument, Gravity benefits from a fully commissioned telescope infrastructure, and employs a sophisticated laser metrology system along the full beam train. However, such an astrometry behind an optical long baseline interferometer suffers from low sensitivity (H~11mag at best with 8.2m UT telescopes), and an extremely small field-of-view (2 arcsec diameter), leading to very low sky coverage, in particular on the extragalactic sky. While Gravity astrometry is focused on precise measurement of multiple-star-systems, MICADO's astrometric mission is to bring this level of plate scale accuracy to measure velocity dispersion in stellar clusters around black holes. If achievable, this would allow us for instance to robustly search for intermediate mass black holes, which are in cosmology the missing link between stellar mass and galaxy mass BHs. In the following, we sketch out what needs to be done to achieve this goal.

## 3. MICADO ASTROMETRY FUNDAMENTAL LIMITS – SCALE: 0.02 ARCSEC

*Detectors:* we plan for standard HgCdTe detector technology, which typically delivers pixel grids at $10^{-3}$ pixel pitch accuracy. We will work with 1.5 and 4mas pixel scales, which slightly oversamples the NIR diffraction limit of the ELT (FWHM/D = 8.5 mas at 1.6 µm). At $10^{-3}$ sub-pixel level, interpixel charge bleeding, and non-identical intra-pixel sensitivity patterns start to play a role (for instance the so-called brighter-fatter effect was recently measured at this level in HgCdTe NIR detectors [2]), so we assume 5 µas as a fundamental limit for MICADO centroiding without further calibration of the detector intra- and inter-pixel behaviour.

*Spatial variation of the image quality* at distances << 1arcsec affect distance measurements at the same level of a few $10^{-3}$ pix, this includes filters, and ADC, which however require separate full field calibrations. Meaning, filters and ADC do introduce noticeably change of the distortion, but being close to the pupil the wavefront errors related to these transmissive optics have very small effects on the PSF shape variation and these very small scales.

*The optical design* of MICADO and MAORY creates a distortion pattern, which to a large extent can be described with a 10 two-dimensional parameter polynomial of up to 3rd order [3]. Given the rectangular shape of the detectors, Legendre polynomials are ideal due to being orthogonal on normalized rectangular support [5]. Mid-spatial frequency errors on surfaces close to the focal plane will give higher-order distortion patterns, but using high-SNR focal plane masks calibrations [7], we can describe those to a level of 5uas in the smallest pixel scale.

Summarizing, these error terms will add up to an accuracy $\sigma_{high.order} \approx 10$µas comparable to the $\sigma_{centroid}$ of an SNR=1000 star. Thus, the typical noise-floor of a MICADO astrometric measurement (ie. a distance measurement involving two stars with uncorrelated $\sigma_{high.order}$ and $\sigma_{centroid}$) will be 20uas, as the quadratic sum of two star's centroiding, which each are limited by $\sigma_{centroid}$ and $\sigma_{high.order}$.

At the smallest scales resolvable, ~10-20mas, these fundamental properties of the imaging measurement limit the achievable local plate scale measurement accuracy to $10^{-3}$.

## 4. DISTORTION DRIFTS START TO CONTRIBUTE AT SPATIAL SCALES 0.02-2 ARCSEC

At larger scales, $\sigma_{centroid}$ and $\sigma_{high.order}$ will not change, which seems to lead to more and more precise star-star distance measurement. However, now the drift of the macroscopic distortions terms start to be relevant. To avoid confusion, we clarify at this point: the typical application of MICADO astrometry will be a multi-epoch stellar proper motion measurement. The key limitation of the astrometric accuracy of such an experiment is not necessarily the distortion itself, but the *temporal stability* of the distortion pattern. As was presented in many observations with the HST, even undersampled PSFs and significant optical distortions can be precisely calibrated to impressive levels of sub-mas astrometric accuracy [8], but this works particularly well only for the optically extremely stable space telescope. We cannot expect a similar stability from ELT and the MAORY / MICADO assembly, which suffer from changing gravity (flexure), wind load, active telescope control, adaptive optics residuals to name a few. While the mean distortion can be measured the limitation is likely to come from understanding the *drift* of the distortions with time, pointing etc. as recently established for the 8m class GeMS MCAO imager [5].

We explored the sensitivity of the geometric distortion to changes in the optical path (mirror position stability, optical interface stability between instrument and telescope, drift residuals of the MCAO loop…). While both MAORY and MICADO act gravity invariant, flexure induced geometric distortions drifts will be most significant in the ELT. We estimated, that after removing the dominating linear terms, the drifting distortions can move stars by up to $\sigma_{3rd.order} \approx 100\mu as$ [3] rms over the entire FoV.

Given that the astrometric field of view has a typical scale length of 20arcsec over which this $\sigma_{3rd.order}$ occurs, at 2arcsec distance levels, this additional error term equals and will start to dominate of $\sigma_{high.order}$ and $\sigma_{centroid}$ and needs to be considered (thus monitored) the larger the distances are. The affine linear terms (scale, rotation, shear are labelled here: $\sigma_{plate.scale}$) typically drift few ten times more than the leading 3rd order distortion terms for optical systems. For the ELT, the plate scale stability during observations is expected as $10^{-4}$, so 2mas over the 20arcsec scale length. Thus for the ELT, distortion drift errors $\sigma_{plate.scale}$ is 20times larger than $\sigma_{3rd.order}$ [3]. The drift of larger than third order distortion terms of the opto-mechanics are negligible in the here discussed context. In a recent experiment with the Gemini GeMS MCAO imager, we saw a similar relative strength of $\sigma_{plate.scale} / \sigma_{3rd.order}$ [5].

This leads to the interesting conclusion, that the expected drift $\sigma_{3rd.order}$ of the low order distortion terms is only relevant for the long-term stability of MICADO astrometry, if the plate scale can be controlled and in particular calibrated to the $10^{-5}$ level. This will be a prime task for the macroscopic astrometric calibration of MICADO images. This can be achieved by three stars in the few 10arcsec scale FoV with (relative) astrometric position accuracy at the 100 µas level.

A potential further relevant astrometric drift term is the astrometric (distortion) residual of the MCAO responding to flexure in the telescope. The MAORY deformable mirrors (DMs) are conjugated to the dominant atmospheric turbulence layers above the ground at Cerro Armazones (about 4km and 15km). The respective wavefront sensors of the telescope will see however a fraction of the optical wavefront error coming from the telescope as well, and send respective correction signals to the DMs. Since DMs and telescope mirrors are not conjugated to each other, a full compensation of such error due to MCAO is impossible, plus the resulting distortions are not directly seen / controlled. The results of on-sky MCAO astrometric experiments with GeMS are encouraging, and seem not to show a major residual here. The detailed simulation and analysis of this error term for ELT/MICADO will be one of our next steps towards calibrating the MICADO astrometry, but is beyond of this status presented in this paper.

## 5. PLATE SCALE CALIBRATION LIMITS ASTROMETRY ON SCALES 2-20ARCSEC

Almost every drift of an optical property leads predominantly to a linear, affine transformation, stretching one scale (like atmospheric refraction, focalplane tilt), two scales (plate scale), and image rotation. Since the telescope is a significant player which does not deliver linear-term accuracy to our requirements, our strategy is two-fold:

### 5.1 Instrumental linear-term control

The MCAO system will naturally stabilize the plate scale precision during the observations (typically a few minutes) to levels $10^{-5}$ when delivering diffraction limited performance with Strehl ratios above 0.2, which is a prerequisite to conduct high-performance astrometric observations. It will deliver PSFs stabilized to about 0.5mas, over a 50arcsec sized FoV. The image derotation is designed to achieve the same over the MICADO field. Achieving a 0.5mas centroid stability over a 50arcsec diameter field translates into an image derotation precision of 2arcsec during the integration [4,6].

Slow drifts of the linear terms of MAORY+MICADO are caused by drifting optical interfaces due to bending of the Nasmyth instrument support platform and deformation of the MICADO instrument support structure, as well as thermal drifts over the entire meter-scale structure. Typical coefficient of thermal expansion (CTE) of steel are $10\text{-}20*10^{-6}$/K, so typical nightly ambient temperature drifts of a few K lead to several 10s of microns on meter scale structures. Also the post-focal DM reference surface positions can drift. These effects can be tracked by the focal plane mask calibration, and will occur at the same time scale of the ELT linear term drifts. While MICADO distortions will co-rotate with the field (and therefore be stable in time), MAORY distortions will not. To ensure that the non-rotating MAORY distortions will not feedback into the science field via the NGS WFS, a careful cross-calibration in daytime is planned.

Our calibration mask will be manufactured with at least $0.5*10^{-5}$ relative accuracy on the linear terms of the reference source grid [7]. The optical design of MICADO and MAORY ensures no significant non-linear distortion on spatial scales smaller than 2 arcsec (Sect. 4). Therefore at smaller scales the imaging is fully described by a three-parameter linear transformation, and we design our reference distortion grid (astrometric calibration mask) with a grid separation of a few 100 mas. This oversampling will allow for a robust estimation of the local scale, as well as for outlier rejection of individual problematic mask holes or centroid estimates.

Summarizing, we expect to stabilize the instrumental linear terms to $10^{-5}$ levels at a time-scale of an hour, and on spatial scales of 2arcsec and larger. The time-scale is given both by typical significant temporal variation of telescope pointing and ambient temperature. On smaller spatial scales we will be limited by other effects (Sect. 3). Once MCAO and MICADO are calibrated to this level, the low order distortion drifts are dominated by telescope and atmosphere. Our analysis of the telescope stability concluded that this large spatial scale low order distortion stability is actually fully dominated by the linear terms (Sect.4). Also residuals from atmospheric refraction and AO corrections are expected to be dominated at the $10^{-5}$ level by linear terms.

To control these remaining large-scale (> 2arcsec) low-order drifts we need:

### 5.2 Sky calibration

We rely on sky calibration of the overall affine distortion of the image. In the typical Galactic astrometry fields (clusters) this can be done with Gaia stars (we expect 0.1-1mas accuracy / 10arcsec scales in 2025+ from the Gaia catalog, these errors are dominated by the proper motion error estimates and lead to a relative precision of $10^{-4..5}$, just in the right range. Since we only seek to estimate three parameters, averaging over more than three stars is quickly giving higher accuracy. We showed this to be an efficient strategy for stellar clusters in our Galaxy [3]. Extragalactic targets will require self-calibration between epochs using the fact that the global stretch or rotation of the object is negligible. There is also the hope, that cross-calibrating the NIR star catalogue of Euclid against Gaia might provide additional reference sources for MICADO near-infrared astrometric imaging. We investigated the possibility to use distant galaxies as long-term astrometric references, but at the here relevant accuracy of 0.1mas, thus 1/100th of the PSF, it is not clear if the inner-galactic substructure is sufficiently distinguishable from PSF residuals.

With our sky calibration strategy, we intend to rely on reference sources within the scientific field, to avoid observational overheads, but also to avoid repointing during astrometric sky calibration. Unlike ultra-stable space telescopes in zero-gravity environments, repointing of the ELT to reference fields, and the regular recollimation (telescope optical quality control) will lead to significant instabilities and drifts of the ELT distortion at the precision scale we aim to calibrate.

## 6. CONCLUSIONS AND NEXT STEPS

Conceptually similar to a recent engineering study of the GeMS astrometric calibration [5], we will use astrometric masks as distortion reference grids in the warm and cold focal planes of the MICADO-MAORY optical system to stabilize the instrument distortion, down to $10^{-5}$ levels.

Requirement (goal) for MICADO is a plate scale accuracy at $10^{-4\,(-5)}$ levels. This restricts 20μas astrometry to 0.2(2)" distances, which is compliant with the science cases. Our calibration strategy aims to keep the relative precision over larger scales, which leads formally to increasing astrometric errors with spatial scale (ie. 200 μas over 20 arcsec) beyond the fundamental limit of 20 μas.

The fundamental astrometric noise limit will be <=20 μas for typical astrometric campaigns which is 5-20x better than current ground- or space based observations. This will be achieved by Nyquist sampling of the diffraction limit PSF with detector pixels, leading to a massive focal plane array of 3x3 4k detectors (150Mpix) to cover the MCAO corrected field of 53arcsec of MICADO.

Breaking the $10^{-5}$ barrier, thus delivering 20 μas over full field 20arcsec distance scales is in principle possible, but not the aim of the first light version of MICADO, and mainly lacking a straight-forward sky calibration scheme (ie. lacking a sufficiently dense grid of sufficiently bright stars, the sky is the current limit here). Going in the other direction, and trying push down the 20 µas as fundamental limit would include both precise calibration of inter- and intra-pixel effects of the detector, and stellar PSFs measured beyond the SNR~1000 levels, which seems prohibitive for an early effective use the ELT time.